\documentclass[%
superscriptaddress,
groupedaddress,
frontmatterverbose, 
preprint,
nofootinbib,
 amsmath,amssymb,
 aps,
]{revtex4-1}

\usepackage{graphicx}
\usepackage{dcolumn}
\usepackage{bm}
\usepackage{hyperref}
\usepackage[mathlines]{lineno}

\begin{document}


\title{Intrinsic energy of Lema{\^{\i}}tre-Tolman-Bondi models and cosmological implications}


\author{Ramon Lapiedra}
 \email{ramon.lapiedra@uv.es}
\affiliation{%
Departament d'Astronomia i Astrof\'{\i}sica, Universitat de
Val\`encia, 46100 Burjassot, Val\`encia, Spain.
}%
\author{Juan Antonio Morales-Lladosa}%
 \email{antonio.morales@uv.es}
\affiliation{%
Departament d'Astronomia i Astrof\'{\i}sica, Universitat de
Val\`encia, 46100 Burjassot, Val\`encia, Spain.
}%
\date{\today}

\begin{abstract} Recently, some Lema{\^{\i}}tre-Tolman-Bondi metrics have been considered as models alternative to the dark energy within the Friedmann-Lema{\^{\i}}tre-Robertson-Walker universes. The vanishing of the intrinsic energy of these metrics is examined since such a vanishing, in the present case and in general, could be interpreted as a necessary condition to consider  the possibility of the quantum creation of a metric. More specifically, this vanishing is examined in the particular case where the Lema{\^{\i}}tre-Tolman-Bondi metrics behave asymptotically like a Friedmann-Lema{\^{\i}}tre-Robertson-Walker universe. Finally, we deal with a particular model ruled out after being confronted with cosmic observations. In a minimal agreement with this negative result, leaving aside an unstable case, the value of the intrinsic energy of this particular model does not vanish and becomes in fact minus infinite.
\end{abstract}


\maketitle



\section{Introduction }
\label{sec:intro}

Out of the particular case where $A' = 0$ (see next section), the Lema{\^{\i}}tre-Tolman-Bondi (LTB) metric family
is the most general family of spherically symmetric metrics in General Relativity, corresponding to a pressure-less matter source \cite{Exact-2003,PlebKra}.

Some of these metrics have been used, in a cosmological context,  to describe large inhomogenous structures and the anisotropies they produce on the cosmic background radiation temperature \cite{AFMS-1993, FSA-1994, FAS-1996} and, more recently,  as models alternative to the dark energy Friedmann-Lema{\^{\i}}tre-Robertson-Walker (FLRW) universes \cite{Celerier,Tomita,Alnes,Enqvist-Mattsson,Enqvist,Hellaby,Wang,Garcia-Bellido,Zumalacarregui,Krasinski-2013} (see also the reviews \cite{BoCeKra-2011,MaNo-20011} on the subject and beyond). The models have been confronted with cosmological observations, and the final conclusion of one of these papers \cite{Zumalacarregui} is that the confrontation has become sufficiently constraining to $``$rule out the whole class of adiabatic LTB models'', by testing a particular LTB model hereafter called the constrained Garc\'{\i}a-Bellido-Haugb\o lle (CGBH) model.

On the other hand, from the beginning of the last seventies, people have speculated on the possibility that the Universe could have arisen
 from a vacuum quantum fluctuation \cite{Albrow,Tryon}, an idea further developed by Vilenkin \cite{Vilenkin-1985}. If this had been the case, we could expect that the energy of our Universe, $P^0$, and also the corresponding linear 3-momentum, $P^i$, and angular 4-momentum, $J^{\alpha\beta}$, would vanish ($i,j, ...=1,2,3$ and $\alpha, \beta, ...=0,1,2,3$). But it is well known that the energy and momenta of a space-time in General Relativity are dramatically dependent on the coordinates used. So, which coordinates must be used in order to calculate the specific two 4-momenta, $P^\alpha$, $J^{\alpha\beta}$, that have to vanish in the case of a \emph{creatable} universe, that is, in the case of a universe arising from a quantum fluctuation? Our answer in \cite{Lapiedra-Morales-2012} is that these specific coordinates have to be \emph{intrinsic} ones, defined as follows:
 
 \begin{enumerate}

\item[(a)] First, they are Gauss coordinates in some region of the considered space-time, covering the 3-space boundary. That is, in this region, the metric components $g_{0\alpha}$ take the values $g_{00}=-1$, $g_{0i}=0$. As is well known, these are coordinates related to free falling synchronized observers (Gauss coordinates have been also selected in \cite{Cooperstock-Dupre} as the ones leading to a sound energy).

\item[(b)] Second, the corresponding linear 3-momentum, $P^i$, and angular 3-momentum, $J^{ij}$, vanish, the last one irrespective of the momentum origin.

\item [(c)] Third, let it be the space-like 3-surface $t=t_0\equiv const.$, with $t$ the time coordinate. We denote this 3-surface by $\Sigma_3$. Then, asymptotically, the 3-space metric, $g_{ij}$, approaches fast enough a manifestly conformally flat metric ($g_{ij}=G^2\delta_{ij}$) when we approach the boundary, $\Sigma_2$, of $\Sigma_3$. Of course, if the space-time is asymptotically Minkowskian, the conformal factor is the unity.

\end{enumerate}

Notice that in Gauss coordinates the time coordinate, $t$, is the proper time of those falling observers. Furthermore, it is a \emph{universal} time. This means that distant readings of this time, that are equal,  correspond to events which are physically simultaneous (see \cite{Landau}, epigraph 84). This could be important in order to define a consistent energy, $P^0$, since, as it is well known and we are going to see, its expression (see next (\ref{energy})) is a 3-volume integral taken in a given time instant $t$. That is, the elementary partial contributions to this integral are taken for the same $t$ value, and it would be a good thing that this common time instant labelled physically simultaneous events. Therefore, it would be good to deal with a proper and \emph{universal} time, i.e., with Gauss coordinates, at least asymptotically,  when calculating what we call here the \emph{intrinsic} $P^0$ energy value: by definition, the one calculated in intrinsic coordinates. As far as the energy, $P^0$, is concerned, the term \emph{intrinsic} refers to the fact that this energy is, first, calculated for coordinates whose corresponding linear and angular 3-momenta, $P^i$ and $J^{ij}$, vanish and, second, since these coordinates are associated with free falling observers (local inertial ones) that, in the present case of LTB metrics, are further co-moving with the source pressure-less matter. Thus, these observers add nothing extra to the considered space-time (compare this situation with, for example, the static observers in a static metric that would have to be  prevented of free falling by some virtual non gravitational action). Therefore, we choose just these coordinates to define the specific $P^0$ energy and momenta, the \emph{intrinsic} 4-momenta, which must vanish if a given metric has to be quantically \emph{creatable}: that is, in our case, if some LTB metric had to be quantically \emph{creatable}.

These intrinsic coordinates can be proved to exist always \cite{Lapiedra-Morales-2012} for each constant value $t=t_0$ and, as mentioned above, the corresponding linear and angular 4-momenta $P^\alpha$ and $J^{\alpha\beta}$ will be called \emph{intrinsic} momenta. In all, the above \emph{creatable} universes would be space-times whose {\em intrinsic} linear and angular 4-momenta, the last one irrespective of the momentum origin, vanish.

We will make explicit the existence of intrinsic coordinates in the particular case of the LTB and FLRW models considered in the present paper. As we will see next, the closed and flat FLRW universes have vanishing \emph{intrinsic} momenta, while the value of the \emph{intrinsic} energy of the open non flat FLRW universe is $-\infty$ \cite{Ferrando}. Thus, in our parlance, the two first ones could be \emph{creatable} universes, but the last one could not be. Furthermore, neither of the perturbed flat FLRW universes in the frame of standard inflation might be \emph{creatable}, while the perturbed closed one could be  \emph{creatable} \cite{Lapiedra-Saez}. Notice that discarding, as we actually do, the physically perturbed flat FLRW model, because of its non {\em creatable} character (and so discarding the unperturbed flat model itself) does not contradict the present standard cosmological model where inflation leads to a FLRW universe with a, perhaps, tremendously small curvature, but not necessarily a strictly vanishing one. 

Let us point out that the starting point for our definition of the intrinsic momenta of a space-time is the Weinberg energy-momentum complex \cite{Weinberg}. There are many complexes present in the literature on the subject, but in \cite{Lapiedra-Morales-2013} we have explained why we have selected, among all them, the one from Weinberg, from which some specific expressions for the energy, $P^0$, the linear 3-momentum, $P^i$, and angular 4-momentum, $J^{\alpha\beta}$, can be deduced. These expressions involve 3-space volume integrals that in the particular case of $P^0$ read
\begin{equation}
P^0=\frac{1}{16\pi}\int \partial_i(\partial_jg_{ij}-\partial_ig)d^3x, \label{energy}
\end{equation}
where the gravitational constant has been taken equal to $1$, $g\equiv \delta_{ij}g_{ij}$ and the summation on repeated indices is performed with the Kronecker $\delta$. This defined $P^0$ coincides (after writing it as a 2-surface integral on the above $\Sigma_2$ boundary) with the well known Arnowitt-Deser-Misner energy \cite{ADM}.

Assuming that an existing universe has to be \emph{creatable} is an attempt at saying something about our Universe all along its existence from just after the big bang. It can also supply us with  criteria for, from the beginning, tentatively discarding,  
as good candidates for good cosmological models all metrics in General Relativity having any non-vanishing intrinsic moment component. This would be the case in the present paper with the observationally discarded CGBH model of the Universe. This discarding could have been suggested from the very beginning because of the physically non \emph{creatable} character of this model. In the present paper, we will characterize the LTB metrics that could be, in principle, {\em creatable} ones, by characterising the ones whose two intrinsic 4-momenta vanish.


\section{The LTB metrics and the general expression for their energy}
\label{sec:2}

The LTB metric class \cite{Exact-2003,PlebKra} can be written as \cite{Enqvist}:
\begin{equation}
ds^2=-dt^2+\frac{A'(t,r)^2}{1-k(r)}dr^2+A^2d\sigma^2\equiv -dt^2+dl^2, \label{metric}
\end{equation}
with $d\sigma^2=d\theta^2+\sin^2{\theta}d\phi^2$ and where $A$ and $k$ are functions of the corresponding arguments satisfying the Einstein field equations, with $A'\equiv \partial_r A$ and $k<1$.  This family of metrics describes the spherical solutions of the Einstein field equations with a pressureless matter source \cite{Exact-2003,PlebKra}. We can see that the coordinates used are adapted to this spherical symmetry. Furthermore, they are Gauss coordinates; that is, for the metric components $g_{0\alpha}$ we have $g_{00}=-1$, $g_{0i}=0$, such that the observers associated to these coordinates are synchronized free radially falling observers co-moving with the pressureless matter source.

We will assume that $A'$ exists and is different from zero everywhere, except perhaps for $r=0$. Note that from (\ref{metric}) we recover the FLRW metrics by putting $A=a(t)r$, $k=\kappa r^2$, with $a$ the corresponding expansion factor and $\kappa$ the curvature index, $\kappa = 1, 0, -1$.

Because of the manifest spherical symmetry of the metric (\ref{metric}), its linear 3-momentum, $P^i$, and its angular 3-momentum, $J^{ij}$, relative to the center $r=0$, vanish (see in detail the case of $J^{ij}$ at the end of Section \ref{sec:5}). Thus, the coordinates used in (\ref{metric}) will be intrinsic coordinates, as defined above, provided that the $r$ coordinate be such that $dl^2$ becomes in the boundary of the corresponding 3-space a manifest conformally flat metric. Then, we are left with the question of whether the energy, $P^0$, calculated in such coordinates, that is, the \emph{intrinsic} energy, vanishes or not in order to conclude if a particular LTB metric could be \emph{creatable} or not.

According to some general expressions given in \cite{Virbhadra}, the expression for the $P^0$ energy of the metric (\ref{metric}) becomes
\begin{equation}
P^0=\frac{1}{2}\lim_{r \rightarrow \infty}\Big[\frac{(A-rA')^2}{r}+\frac{krA'^2}{1-k}\Big], \label{energy bis}
\end{equation}
where we have put both the gravitational constant and the speed of light equal to $1$.

To obtain this expression one must transform the 3-space integral giving $P^0$ in (\ref{energy}) into a 2-surface integral on the boundary of this 3-space by applying Gauss theorem. To apply this theorem we need that the metric be regular enough: the 3-space derivatives of the 3-space metric must be continuous.%
\footnote{If there are no intrinsic singularities in the integration 3-volume, there always exist coordinates in which the Gauss theorem can be applied, provided that we assume, as it is always done, that the differentiable manifold of the General Relativity is ${\cal{C}}^2$-class by pieces. Nevertheless, in the present work, we only have to use intrinsic coordinates. Then, the ${\cal{C}}^2$-class by pieces character could not be fulfilled if we are restricted to only use these intrinsic coordinates. Therefore, for intrinsic coordinates, the above regularity condition in order to apply Gauss theorem should be verified in each case for the particular metric used. \label{Gauss}}
Thus, we will assume not only that, except for $r=0$, $A'$ in (\ref{metric}) exists everywhere and is different from zero, but further than $A''$ is continuous too, and that there is no intrinsic singularity at $r=0$.

As remarked above, since we want this energy $P^0$ to be an \emph{intrinsic} one, we must use in (\ref{energy bis}) an $r$ coordinate such that $dl^2$ in (\ref{metric}) 
can be asymptotically conformally flat in a manifest way. This is what we are going to discuss in the next section.


\section{LTB metrics in asymptotic conformally flat coordinates}
\label{sec:3}

Let us make a transformation of the radial coordinate $r$ in (\ref{metric}), going to a new radial coordinate $\rho$, trough a time independent function, that is $r \rightarrow \rho=\rho (r)$. We will chose this function such that the 3-space metric, $dl^2$, becomes in the new radial coordinate asymptotically conformally flat in a manifest way for any constant time $t_0$. That is,
\begin{eqnarray} \nonumber
dl^2(t=t_0, \rho \rightarrow \rho_b) & \simeq & G^2(t_0, \rho_b, \rho)\delta_{ij}d\rho^id\rho^j \\ 
& = & G^2(d\rho^2+\rho^2d\sigma^2), \label{conf-metric}
\end{eqnarray}
where $\rho^i$ is such that $\delta_{ij}\rho^i\rho^j=\rho^2$ and where $\rho=\rho_b\equiv \rho_b(t_0)$ is the equation of the radial boundary, $\Sigma_2$, of the space-time given by the metric $ds^2$ in (\ref{metric}) at $t_0$. We know such a radial coordinate to exist since in \cite{Ferrando,Lapiedra-Morales-2012} it has been proved on general grounds for every such constant time $t=t_0$. This existence is obvious in the particular case of the LTB metrics which behave asymptotically like FLRW universes (see Sec. \ref{sec:4}).  

Then, by simply comparing (\ref{conf-metric}) with (\ref{metric}), we obtain
\begin{equation}
\frac{A'(t_0, \rho \rightarrow \rho_b)^2}{1-k(\rho \rightarrow \rho_b)}\simeq G^2(t_0,\rho \rightarrow \rho_b),  \label{G-relation}
\end{equation}
\begin{equation}
\rho_b^2G^2(t_0,\rho \rightarrow \rho_b)\simeq A^2(t_0,\rho \rightarrow \rho_b). \label{G-relation2}
\end{equation}

We will write these two equations more compactly by putting
\begin{equation}
\frac{A'^2}{1-k}\simeq G^2, \,\, A^2 \simeq \rho^2G^2. \label{G-relation3}
\end{equation}

These asymptotic equations give trivially:
\begin{equation}
k \simeq 1-\rho^2 \frac{A'^2}{A^2}, \label{k-A relation}
\end{equation}
which, since $k$ only depends on $\rho$, means that asymptotically the function $\rho^2A'^2/A^2$ does not depend on $t_0$.

Then, let us calculate the {\em intrinsic} energy $P^0$ of any LTB metric that is $P^0$ calculated in coordinates such that we have  (\ref{conf-metric}). 
Since we have assumed at the beginning of the Section \ref{sec:2} that our LTB metrics are regular enough, we can, using Gauss theorem, write (\ref{energy}) as the 2-surface integral on the corresponding boundary, 
\begin{equation}
P^0=\frac{1}{16\pi}\lim_{\rho \to \rho_b}\int(\partial_jg_{ij}-\partial_ig)n^i\rho^2\sin \theta d\theta d\phi, \label{surface-energy}
\end{equation}
with $n^i\equiv \rho^i/\rho$.

Since now we have asymptotically $dl^2 \simeq G^2(t_0,\rho)\delta_{ij}dx^idx^j$, $P^0$ becomes
\begin{equation}
P^0=-\frac{1}{8\pi}\lim_{\rho \to \rho_b}\int\partial_iG^2 n^i\rho^2\sin \theta d\theta d\phi=-\frac{1}{2}\lim_{\rho \to \rho_b}(\rho^2\partial_{\rho}G^2), \label{surface-energy2}
\end{equation}
or what is equivalent, according to the second equation of (\ref{G-relation3}), 
\begin{equation}
P^0 = \frac{1}{2} \lim_{\rho \to \rho_b} (2 \rho^{-3} A^2 - \partial_\rho A^2).
\end{equation}

Equalizing  to zero this $P^0$ expression allows us to fully characterize the family of LTB metrics whose two 4-momenta vanish for any constant time $t_0$. For instance, if $\rho_b=  {\infty}$,  then $G^2\sim \frac{1}{\rho^p}$  with $p>1$ gives $P^0 = 0$. This situation describes all the corresponding creatable LTB universes when $\rho_b = \infty$, although there is still another particular case in which $P^0 = 0$: i. e., the one in which $G$ does not depend on $\rho$.


\section{The LTB metrics behaving asymptotically as FLRW universes} 
\label{sec:4}

Imagine that function $A$ factorizes as we approach the boundary, $\Sigma_2$, of the 3-space. That is, we have near $\Sigma_2$,
\begin{equation}
A(t,\rho)\simeq a(t)f(\rho). \label{factorizing}
\end{equation}

Having in mind this factorization, the Einstein field equations (see for example {\cite{Enqvist})
\begin{equation}
\dot A^2+2A\ddot{A}+k=0, \label{A equation}
\end{equation}
\begin{equation}
2 \frac{\ddot{A}}{A}+\frac{\ddot{A'}}{A'}=-4\pi \rho_M,   \label{A equation 2}
\end{equation}
become asymptotically
\begin{equation}
\dot a^2+2a\ddot a=  \rho^2 \frac{f'^2}{f^4} -\frac{1}{f^2} \label{a-f function}
\end{equation}
and
\begin{equation}
\frac{\ddot a}{a}=-\frac{4}{3}\pi \rho_M(t), \label{a function}
\end{equation}
respectively, with $\rho_M(t)=\rho_M(t,\rho \rightarrow \rho_b)$, and $\rho_M(t,\rho)$ the matter density.

From (\ref{a-f function}) we obtain
\begin{equation}
{\dot a}^2+2a {\ddot a}=C,  \quad  \rho^2 \frac{f'^2}{f^4} -\frac{1}{f^2} = C, \label{a and f functions}
\end{equation}
with $C$ an arbitrary constant.

Equation (\ref{a function}) and the first equation of (\ref{a and f functions}) are equivalent to the two dynamical cosmic equations for the expansion factor, $a(t)$, of a FLRW with $-C$ curvature, i.e., equivalent to (\ref{a function}) jointly with
\begin{equation}
\Big (\frac{\dot a}{a}\Big )^2=\frac{8\pi\rho_M}{3}+\frac{C}{a^2}. \label{cosmic eq}
\end{equation}

As far as the second equation of (\ref{a and f functions}) is concerned, its general solution is 
\begin{equation}\label{gensol}
f= \frac{2 B \rho}{1- B^2 C\rho^2}, 
\end{equation}
with $B$ an arbitrary constant. This gives for $G$ the expression:
\begin{equation} \label{Gfunction}
G^2(t,\rho)=a^2(t) \frac{4B^2}{(1- B^2 C \rho^2)^{2}}.
\end{equation}
Then, let us change to the new radial coordinate $\tilde{\rho} = - \frac{2}{BC\rho}$. We obtain for the asymptotic value of this $d s^{2}$ metric
\begin{equation}
ds^2=-dt^2+a^2(t)\frac{\delta_{ij} d \tilde{\rho}^i d\tilde{\rho}^j}{(1-C \tilde{\rho}^2/4)^2}. \label{conformetric}
\end{equation}

That is, all LTB metrics satisfying (\ref{factorizing}) reduce asymptotically to a FLRW universe: close, open or flat according to whether  it is 
$C<0$,  $C>0$, or $C=0$, respectively.  Actually, from (\ref{conformetric}),  we obtain the corresponding standard form
\begin{equation}
ds^2=-dt^2+a^2(t)\Big(\frac{dr^2}{1+Cr^2}+r^2d\sigma^2\Big), \label{FLRW-metric}
\end{equation}
by making the coordinate transformation
\begin{equation}
r=\frac{\tilde{\rho}}{1- C \tilde{\rho}^2/4}. \label{r-rho relation}
\end{equation}

According to previously obtained results (see \cite{Ferrando} for example), the {\em intrinsic} energy, $P^0$, of a closed or flat FLRW universe vanishes, while it gets a $- \infty$ value for the open non flat one. Then, it becomes obvious that $P^0 = 0$ for any LTB metric behaving asymptotically like a closed FLRW universe, and $P^0 = - \infty$ when it behaves asymptotically like an open non flat one. But what about the case where the LTB metric approaches asymptotically a flat FLRW universe? We deal with this question in the next section.


\section{The particular case of the LTB metrics asymptotically behaving like a flat FLRW universe}
\label{sec:5}

In references \cite{Garcia-Bellido,Zumalacarregui}, the corresponding authors explore the possibility that we live close to the center of a large void, i.e. close to the center of a suitable LTB model, as an alternative to the prevailing interpretation of the Universe acceleration in terms of a $\Lambda$CDM model with a dominant dark energy component. They confront this possibility with a series of cosmological observations through two versions, the flat and the open ones, of the CGBH model cited above, the first (second) version becoming asymptotically a flat (an open non flat) FLRW universe without cosmological constant. The CGBH model, in its two versions, is ruled out as a result of the confrontation.

Because of the asymptotic behaviour of the open non flat version of this CGBH model, this intrinsic energy $P^0$ has to be $- \infty$. This result is in minimal accordance with the reported ruling out of this version by observational reasons (remember that, in our frame, universes with at least one of the components of their intrinsic 4-momenta different from zero would not be a good candidate to be quantically creatable).  Of course, getting $P^0 = - \infty$ from this asymptotic behaviour assumes that the metric of this open non flat version is everywhere regular enough in order to apply the Gauss theorem. The same question will be raised below, in the present section, for the flat version. We will postpone to the end of this section the proof that, in both cases, the theorem becomes applicable.

But, what about the asymptotic flat version of the CGBH model? In the present section, we will see that in this case the {\em intrinsic} energy $P^0$ vanishes or not depending on how fast the flat version of the CGBH model approaches asymptotically a flat FLRW universe.

Then, let us compare the LTB metric (\ref{metric}) with its flat FLRW universe limit for $r \to \infty$. Having in mind the expression
$ds^2=-dt^2+a(t)^2(dr^2+r^2 d\sigma^2)$ for this flat FLRW metric, we easily obtain
\begin{equation}
A(t, r \to \infty)\simeq a(t)r,\quad \quad  \lim_{r \rightarrow \infty}k(r)=0. \label{limits}
\end{equation}
Notice that, because of this asymptotic character, the coordinates in (\ref{metric}) are asymptotic conformally flat coordinates for $dl^2$. Consequently, they are intrinsic coordinates (they fulfill the three conditions $\rm{(a)-(c)}$ defining the notion of intrinsic coordinates in the Introduction) and the corresponding $P^0$ energy we are going to calculate will be the \emph{intrinsic} energy.

Because of (\ref{limits}), the partial contribution to $P^0$, in (\ref{energy bis}), coming from the term containing the $k$ function is
\begin{equation}
\frac{1}{2}\lim_{r\rightarrow \infty}\frac{rkA'^2}{1-k}=\frac{a^2}{2}\lim_{r\rightarrow \infty}\frac{rk}{1-k}=\frac{a^2}{2}\lim_{r\rightarrow \infty}(rk), \label{contribution}
\end{equation}
whose actual value depends on how fast $k$ vanishes when $r\to\infty$. Then, in accordance with \cite{Garcia-Bellido}, let us define the function $\Omega_M(r)$, that generalizes the matter cosmic parameter, $\Omega$, of the FLRW cosmology, by writing $k(r)$ like
\begin{equation}
k\equiv \dot A_0^2(\Omega_M-1), \label{definition Omega}
\end{equation}
with $\dot A_0\equiv \partial_t A(t=t_0,r)$ and $t_0$ the cosmic present time.

The value of this function for the asymptotic flat version of the CGBH model is \cite{Garcia-Bellido}
\begin{equation}
\Omega_M=\Omega_{out}+(\Omega_{in}- \Omega_{out})\frac{1-\tanh[(r-r_0)/2\Delta r]}{1+\tanh (r_0/2\Delta r)}, \label{expression Omega}
\end{equation}
with $\Omega_{out}=1$ ($\Omega_{out}<1$, for the open non flat case), where $\Omega_{in}$, $r_0$,  and $\Delta r$ are parameters to be fitted by cosmological observations. In particular, $r_0$ characterizes the void size, near whose center we are assumed to be placed, and $\Delta r$ the transition to uniformity.

From (\ref{expression Omega}), we easily obtain
\begin{equation}
\Omega_M(r \gg \Delta r)-1\simeq\lambda e^{-r/\Delta r}, \label{asymptotic Omega}
\end{equation}
with
\begin{equation}
\lambda\equiv \frac{2e^{r_0/\Delta r}}{1+\tanh (r_0/2\Delta r)}(\Omega_{in} - 1). \label{lambda}
\end{equation}
Then, from (\ref{definition Omega}) and the first equation of (\ref{limits}) we have
\begin{equation}
k(r \gg \Delta r)\simeq \lambda {\dot{a}}_0^2r^2e^{-r/\Delta r}. \label{asymptotic k}
\end{equation}

According to (\ref{contribution}), this asymptotic behavior of $k$ means that the contribution to $P^0$ in (\ref{energy bis})
from the term involving $k$ vanishes. Thus, in order to calculate $P^0$ for the asymptotic flat version of the  CGBH model, we are left with the remaining contribution from the term dealing with the function $A$. To calculate this contribution we have to make explicit how fast $A$ approaches $a(t)r$ (see the first equation of (\ref{limits})) when $r \to \infty$. The easiest way to make this is to use one of the Einstein field equations (\ref{A equation}) and (\ref{A equation 2})  for the LTB metrics (\ref{metric}).

More precisely, we will consider Eq. (\ref{A equation}) for $r \to \infty$ where,  as we have just seen,  $k$ behaves like $k\sim r^2e^{-r/\Delta r}$, jointly with the parametric $A$ value for the asymptotically  flat CGBH model \cite{Garcia-Bellido,Zumalacarregui}, i.e.,
\begin{equation}
A(t,r)=\frac{\Omega_M}{2(1-\Omega_M)}(\cosh \eta-1)r, \label{A-parameter-1}
\end{equation}
\begin{equation}
H_0(r)t=\frac{\Omega_M}{2(1-\Omega_M)^{3/2}}(\sinh \eta-\eta), \label{A-parameter-2}
\end{equation}
with $\eta$ a positive real parameter, $\eta \in (0, + \infty)$, and
\begin{equation}
H_0(r)=\frac{3 H_0}{2 (1-\Omega_M)}\Big[1-\frac{\Omega_M}{\sqrt{1-\Omega_M}}\sinh^{-1}\sqrt{\frac{1-\Omega_M}{\Omega_M}}\Big],
\label{function-H0}
\end{equation}
with the inserted factor $3/2$ allowing us to obtain the Hubble constant $H_0$, for $H_0(r \to \infty)$. Finally, $\Omega_M$ is given by (\ref{expression Omega}) with $\Omega_{out}=1$.

Then, in accordance with (\ref{limits}), let us write the asymptotic form of $A$ as
\begin{equation}
A(t,r \to \infty)=a(t)r[1+\epsilon(t,r)], \label{corrected-A}
\end{equation}
where $\epsilon$ is a function of $t$ and $r$ such that $\lim_{r \rightarrow \infty}\frac{\epsilon}{r^p}=0$ for any $t$ value and for any $p>1$. This asymptotic expression for $A$ would guarantee the vanishing of the corresponding contribution to the \emph{intrinsic} value of $P^0$. In the Appendix we show that $A$ has actually such an asymptotic behavior since we obtain
\begin{equation}
\epsilon(t,r \to \infty)\sim e^{-r/\Delta r}. \label{epsilon}
\end{equation}

Then, the resulting asymptotic form for $A$ when $r \to \infty$ leads to
\begin{equation}
\lim_{r \rightarrow \infty}\frac{(A-rA')^2}{r}=0 \label{first-contribution}
\end{equation}
and so to the final vanishing of the \emph{intrinsic} energy of the asymptotically flat CGBH model.

It seems, then, that the asymptotically flat CGBH cosmological model could be quantically  \emph{creatable}. However, we are going to see in Section \ref{sec:6} that this vanishing of $P^0$ is a dramatically non stable result. We will account for this limitation by saying that the model is \emph{physically} non \emph{creatable}. 

But, according to the general considerations on the applicability of Gauss theorem, made at the end of Section \ref{sec:2} (including footnote \ref{Gauss}), we will show before going into Sec. VI, that the metric of both versions of the CGBH model is regular enough to allow for the application of this theorem:
we need to apply the theorem to go from (\ref{energy}) to (\ref{surface-energy}) in order to calculate $P^0$. In our case, a sufficient condition for the theorem is that the first $r$ derivatives of the  3-space metric components of the LTB metric (\ref{metric}) be continuous functions everywhere. In the particular case of the CGBH model, this means that the second $r$ derivatives of $A(t,r)$ given by (\ref{A-parameter-1}), (\ref{A-parameter-2}), (\ref{function-H0}),  and (\ref{expression Omega}),  is continuous, and the same for the first $r$ derivative of $k(r)$
given \cite{Garcia-Bellido} by $k(r) = r^2 H_0^2(r)[\Omega_M(r)-1]$.

But notice that, to begin with, the function $\Omega_M(r)$ is infinitely derivable since it is essentially $\tanh r$. On the other hand, the first $r$ derivative of the function $\sinh^{-1}\sqrt{(1-\Omega_M) \Omega_M^{-1}}$ in (\ref{function-H0}), when calculated, gives the value $- (1/2) \Omega'_M \Omega_M^{-1} (1- \Omega_M)^{-1/2}$. Then, it is straightforward to see that the function $H_0(r)$ given by (\ref{function-H0}) has a continuous first $r$ derivative everywhere, out perhaps of $\Omega_M = 0$ and $\Omega_M = 1$. But, by definition $\Omega_M \neq 0$ (notice \cite{Zumalacarregui} that $\Omega_M \equiv \bar{\rho}(r)/\rho_c$, where $\rho_c \equiv 3H_0(r)/8 \pi$ is the critical density and 
$\bar{\rho}(r) = (1/V) \int_0^r 4 \pi r'^2 \rho({r', t_0) dr'}$, with $V=4 \pi r^3/3$). Further, there is no physical singularity for $\Omega_M \to 1$, since in this case we simply recover asymptotically the corresponding flat FLRW limit. Then, in $H_0(r)$ given by (\ref{function-H0}), we can  leave out the particular $\Omega_M$ values $\Omega_M=0$, $\Omega_M=1$, so that $H'(r)$ is continuous everywhere. In all, $k'(r)$ is a continuous function as we wanted to prove.

Now, let us prove the continuity of $A''(t, r)$. In order to do this, let us write Eq. (\ref{A-parameter-2}) as
\begin{equation}
\sinh \eta-\eta = 2 \, t \,  H_0(r) \Omega_M^{-1} (1-\Omega_M)^{3/2} \equiv t \chi (r),  \label{A-parameter-2b}
\end{equation}
where $\chi (r) \equiv 2 H_0(r) \Omega_M^{-1} (1-\Omega_M)^{3/2}$.

But the function $\sinh \eta-\eta$ of $\eta$ is infinitely $r$ derivable, monotonously increasing from $0$, for $\eta \to 0$, to $+ \infty$,  for $\eta \to +\infty$. Then, the corresponding inverse  function of $r$, $\eta = f(t \chi)$,  is unique and infinitely derivable.

Thus, let us calculate $A'(t, r)$ from (\ref{A-parameter-1}):
\begin{equation}
A'(t,r)=\frac{1}{2} \Big\{\Big[\frac{r \Omega_M}{1 - \Omega_M}\Big]' (\cosh f - 1) +  \frac{r \Omega_M f'}{1 - \Omega_M}
\sinh f  \Big\}. \label{A-parameter-prima}
\end{equation}
where, $\cosh f$ and $\sinh f$ are infinitely derivable functions of $r$ and the same for $f' \equiv \chi' \partial_\chi f$ according to the above reasoning. Then, by a mere inspection of 
(\ref{A-parameter-prima}) we can see that (out of $\Omega_M=0$ and $\Omega_M=1$, as it is due) $A''$ exists and is continuous.

Therefore, the sufficient regularity conditions to apply Gauss theorem to (\ref{energy}) are fulfilled.

Before finishing the present section, let us see in detail how the explicit spherical symmetry of a 4-metric in Gauss coordinates leads to the vanishing of the corresponding $J^{ij}$, as announced 
at the beginning of Section \ref{sec:2}. Notice that, because of this symmetry, when using rectilinear coordinates at the space infinity $r \to \infty$, the 3-space metric, $g_{ij}$, has the asymptotic form:
\begin{equation}
g_{ij} = \alpha(t, r) \delta_{ij} + \beta(t,r) n_i n_j, \,  n_i = \frac{x_i}{r}, \,  r= \sqrt{x_i x_i}, \label{3-metric-asymptotic}
\end{equation}
where $\alpha(t, r)$ and $\beta(t, r)$ are two functions of $t$ and $r$.

On the other hand, and similar to expression (\ref{surface-energy}), starting from the Weinberg complex \cite{Weinberg}, we obtain for $J^{ij}$ referred to the angular momentum origin $r=0$, the following general expression as a 2-surface integral on the boundary $r \to \infty$ of the 3-space
\begin{equation}
J^{ij} = \frac{1}{16 \pi} \lim_{r\to\infty} \Big[r^3 \int (n_j \dot{g}_{ki} - n_i \dot{g}_{kj}) n_k \sin \theta d\theta d\phi \Big]. \label{3-momentum-int}
\end{equation}

Of course, we need to apply the Gauss theorem to the corresponding 3-volume integral in order to obtain (\ref{3-momentum-int}). In our case, having just been proved above that $k(r)$ is regular enough,  this requires as a sufficient condition that $\dot{A}'$ be a continuous function of $r$. But from (\ref{A-parameter-prima}) we obtain 
\begin{equation}
\dot{A}'(t,r)=\frac{r  \Omega_M \dot{f} \sinh f }{2 (1 - \Omega_M)} = \frac{r  \Omega_M \chi \sinh f}{2 (1 - \Omega_M)} \frac{df}{d(t \chi)}, \label{A-parameter-puntprima}
\end{equation}
which, for reasons similar to the ones explained above to conclude the everywhere continuity of $A''$, is an everywhere continuous function of $r$. Therefore, we can write (\ref{3-momentum-int}) for $J^{ij}$ referred to the angular 3-momentum origin $r=0$. Finally, we can substitute (\ref{3-metric-asymptotic}) in (\ref{3-momentum-int}) and obtain $J^{ij} = 0$ since it is obvious that now the integrand vanishes identically. 

But, what if we shift the angular 3-momentum  origin from the above value $r=0$ to $r = \sqrt{a_i a_i} \equiv a$, where $a_i$ are the components of a constant 3-vector? We will have in an evident notation
\begin{equation}
\bar{J}^{ij}= \frac{1}{16 \pi} \lim_{r\to\infty} \Big[r^2 \int (\bar{x}_j \dot{g}_{ki} - \bar{x}_i \dot{g}_{kj}) n_k \sin \theta d\theta d\phi \Big] \label{3-momentum-int-a}
\end{equation}
where $\bar{x}_j \equiv x_j + a_j$.

A sufficient  condition  to have $\bar{J}^{ij} = 0$ irrespective of the chosen angular 3-momentum origin, that is, irrespective of the constant value $a_i$, is to have:
\begin{equation}
I_i \equiv  \int \dot{g}_{ki} n_k \sin \theta d\theta d\phi = 0. \label{int-1}
\end{equation}

But, according  to  (\ref{3-metric-asymptotic}), this integral actually becomes
\begin{equation}
I_i =  (\dot{\alpha} + \dot{\beta})\int n_i \sin \theta d\theta d\phi = 0,  \label{int-2} 
\end{equation}
since identically $\int n_i \sin \theta d\theta d\phi = 0$. 

Thus, for any manifestly spherically symmetric 4-metric in Gauss coordinates, $J^{ij}$ vanish identically irrespective of the chosen angular 3-momentum origin.

For completeness we can report that $J^{0i}$ vanish too, irrespective of the chosen origin of angular 4-momenta simply because, in the present case $P^0=0$ and $P^{i}=0$.

This double vanishing happens too for the CGBH model in intrinsic coordinates, since in such a particular case, aside from being $P^0=0$ and $P^{i}=0$, Eq. (\ref{3-metric-asymptotic}) reduces to $g_{ij} = \alpha(t, r) \delta_{ij}$, as it must according to the intrinsic coordinate definition (see point {\rm c} of this definition in the Introduction). 

The reader can see that asymptotically $g_{ij} = \alpha(t, r) \delta_{ij}$ for the open non flat version of the CGBH model by consulting (\ref{conformetric}) for $C>0$. For the flat version, this asymptotic behaviour, $g_{ij} = \alpha \delta_{ij}$, becomes obvious by having in mind (\ref{asymptotic k}), (\ref{corrected-A}),  and (\ref{epsilon}).


\section{The instability of the energy vanishing for a LTB metric behaving asymptotically as a flat FLRW universe: discussion and conclusion}
\label{sec:6}

Let us come back to the open non flat FLRW metric in the form (\ref{conformetric}) with $C$ normalized to the corresponding value $C=1$. The coordinates  used are then intrinsic ones and, according to (\ref{surface-energy2}), the intrinsic value of $P^0$ becomes, 
\begin{equation}
P^0=-\frac{a^2}{2}  \lim_{\rho \to 2}\Big[\rho^2\frac{d}{d\rho}(1-\rho^2/4)^{-2}\Big]=-\infty, \label{infinite-energy}
\end{equation}
since in this case it is $\rho_b = 2$.

Therefore, $P^0$, whose value for the asymptotically flat CGBH model (Section \ref{sec:5}) was zero, jumps to a minus infinite value, $P^0=-\infty$, when an elementary shift of the $C$ constant from its original value is performed, or what is the same, when a shift from $\Omega_{out}=1$ (see (\ref{expression Omega})) to the new value $\Omega_{out}<1$ is performed as close to $1$ as we want.

Thus, in the frame of a hypothetical quantum creation of a universe (this universe being constrained to have vanishing energy in accordance with, for example, authors like Tryon \cite{Tryon} or Vilenkin \cite{Vilenkin-1985}), the resulting vanishing of $P^0$ for this asymptotically flat CGBH model could be considered as an unstable result, and so the quantum creation of this asymptotically flat model could perhaps have a vanishing probability. All this is, of course, compatible with some inflationary process leading to a universe extremely (but not exactly) flat, as is the case in the present cosmological standard model. Then, though in our frame this flat model could be strictly speaking a \emph{creatable} one, we can consider and denote it, jointly with the asymptotically non flat open CGBH model, as physically non \emph{creatable} universes. All this could be seen in accordance with the fact that this model is ruled out by its confrontation with cosmological observations \cite{Garcia-Bellido, Zumalacarregui}. In conclusion: had this confrontation not still taken place, in view of this physical non \emph{creatable} character, we could have predicted a subsequent negative result, not as a true prediction but as some plausible suggestion.

Furthermore, assuming a negative value for the constant $C$ in (\ref{conformetric}) and normalizing to $C=-1$ we obtain the closed FLRW metric in 3-space conformal flat coordinates:
\begin{equation}
ds^2=-dt^2+a^2(t)\frac{\delta_{ij}d\rho^i d \rho^j}{(1+\rho^2/4)^2}. \label{closedmetric}
\end{equation}

The boundary limit $\rho_b$ is now $\rho_b=\infty$. Then, using the expression (\ref{surface-energy2}), we straightforwardly obtain $P^0=0$; that is, the LTB metric approaching asymptotically the closed FLRW model could be \emph{creatable}. Thus, we could ask if this $``$closed'' LTB model would not be better entitled than the CGBH one to be observationally tested as a model of the void universes considered in the present paper. Actually, aside with other models, the  $``$closed'' LTB model  has been tested in \cite{Wang}, without fully conclusive results. But, obviously, the possible \emph{creatable} character of this model, by itself, could not avoid that finally the model was ruled out by observations. In fact, as stated in \cite{Wang}, $``$in practice we could always approximate the correct answer by setting $\Omega_k$ to a small nonzero value''. 

Nevertheless, according to \cite{Phil-Buy} (cf. \cite{Clarkson-Regis} and references therein), it seems that the kinematic Sunyaev-Zel'dovich effect puts further severe limitations on the viability of simple LTB void models, irrespective of their asymptotic (closed, open or flat) character and the constraining requirement of a homogeneous universe age.


\begin{acknowledgments}

This work was supported by the Spanish 
``Ministerio de  Econom\'{\i}a y Competitividad'', MICINN-FEDER project FIS2012-33582. 

\end{acknowledgments}


\appendix
\section{Calculating the $\epsilon(t,r)$ function}
\label{ap-A}

We start from (\ref{A equation}), that is, $\dot A^2+2A\ddot{A}+k=0$, and from (\ref{asymptotic k}), that is, $k(r\gg\Delta r) \simeq \lambda {\dot{a_0}}^2r^2e^{-r/\Delta r}$. Then we write $A$ as $A=a(t)r[1+\epsilon(t,r)]$ with $a(t)= {(t/t_0)}^{2/3}$. After an elementary calculation, neglecting quadratic terms in $\epsilon$, we find
\begin{equation}
\frac{1}{r^2}(\dot A^2+2A\ddot{A})\simeq ({\dot{a}}^2+2a\ddot{a})(1+2\epsilon)+2a(a\ddot{\epsilon}+3\dot{a}\dot{\epsilon}).
\label{left hand side}
\end{equation}
But we have ${\dot{a}}^2+2a\ddot{a}=0$. Then, Eq. (\ref{A equation}) becomes for large values of $r$
\begin{equation}
2a(a\ddot{\epsilon}+3\dot{a}\dot{\epsilon})\simeq -{\dot{a}_0}^2 \lambda{e^{-r/\Delta r}}, \label{epsilon-eq}
\end{equation}
which in accordance with $a(t)={(t/t_0)}^{2/3}$ can be written as
\begin{equation}
(a^3\dot{\epsilon})\dot{}\simeq -\frac{2}{9{t_0}^2}\lambda e^{-r / \Delta r}, \label{epsilon-eq2}
\end{equation}
whose general solution for large values of $r$ ($r \gg \Delta r$) is
\begin{equation}
\epsilon(t,r)=g(r)\alpha(t)+h(r), \label{epsilon-gral-solution}
\end{equation}
with
\begin{equation}
g(r)=-\frac{2}{9{t_0}^2}\lambda e^{-r / \Delta r},\label{f-function}
\end{equation}
where $h(r)$ is an arbitrary function, and $\alpha(t)$ the general solution of
\begin{equation}
(a^3\dot{\alpha})\dot{}=a, \label{alpha-eq}
\end{equation}
that is to say,
\begin{equation}
\alpha=\frac{9}{10}{t_0}^{4/3}t^{2/3}+\mu {t_0}^2t^{-1}+\nu, \label{alpha solution}
\end{equation}
with $\mu$ and $\nu$ two arbitrary constants.

Substituting this expression of $\alpha$ in (\ref{epsilon-gral-solution}), we obtain, for large values of $r$ but for any time,
\begin{equation}
\epsilon(t,r) = -\Big[\frac{1}{5}a+\frac{2}{9}(\frac{\mu}{t}+\frac{\nu}{{t_0}^2})\Big]\lambda e^{-r / \Delta r}+h(r). \label{explicit-solution}
\end{equation}

Then, in order to fix both arbitrary constants, $\mu$ and $\nu$, and the arbitrary function $h(r)$, let us come back to the CGBH model, more precisely, to Eqs. (\ref{A-parameter-1}) and (\ref{A-parameter-2}). For small $\eta$ values, that is, for small $t$ values, we obtain
\begin{equation}
A(t,r)\simeq \Big(\frac{3}{2}\Big)^{2/3}{\Omega^{1/3}_M(r)}{H^{2/3}_0(r)} \, r \, t^{2/3}, \label{A-small t}
\end{equation}
where we can substitute $\Omega_M$ by its asymptotic expression from (\ref{asymptotic Omega}).

Furthermore, having in mind (\ref{expression Omega}) with $\Omega_{out}=1$ and (\ref{function-H0}), after an elementary calculation, we obtain the following asymptotic value:
\begin{equation}
H(r \gg \Delta r)\simeq H_0(1-\frac{1}{5}\lambda \, e^{-r / \Delta r}). \label{H}
\end{equation}

Carrying this expression to (\ref{A-small t}) we obtain for large values of $r$ and small values of $t$
\begin{equation}
A(t,r)\simeq ar\Big[1+\frac{1}{5}(1-a)\lambda \, e^{-r/\Delta r}\Big ]. \label{asymptotic A}
\end{equation}

Finally, through the relation $A=ar(1+\epsilon)$, let us compare this approximated expression of $A(t,r)$ with $\epsilon(t,r)$, given by (\ref{explicit-solution}). We obtain for large values of $r$, but for any time,
\begin{equation}
\epsilon\simeq \frac{1}{5}\lambda (1-a)e^{-r / \Delta r}, \label{final solution}
\end{equation}
which is in accordance with Eq. (\ref{epsilon}).


\bibliography{apssamp}

\end{document}